# Electron Energy Losses Near Pulsar Polar Caps

Steven J. Sturner[1]
E. O. Hulburt Center for Space Research, Code 7653,
Naval Research Laboratory, Washington, DC 20375-5352

## ABSTRACT

We present results of a model for the energetics of electrons accelerated by the large electric fields generated by a rotating highly magnetized neutron star. The energy loss mechanisms we consider in our calculations include magnetic Compton scattering of thermal x-ray photons, triplet pair production, and curvature radiation emission. The electron acceleration mechanism is assumed to operate only to a height above the polar cap approximately equal to the polar cap radius. We find several interesting results. First, magnetic Compton scattering is the dominant energy loss process when the electron Lorentz factors are $<$ few$\times 10^6$ for typical gamma-ray pulsar magnetic fields and surface temperatures measured by *ROSAT*. The amount of energy converted to photons by accelerated electrons ranges from $\sim 10\%$ to $\sim 100\%$ of $\gamma_o m_e c^2$ where $\gamma_o$ is the maximum Lorentz factor an electron can attain with no radiative losses. We also find that if $B > 10^{13}$ G and $T > 3 \times 10^6$ K, the Lorentz factors of the electrons can be limited to values $\lesssim 10^3$ assuming values for the size of the neutron star thermal polar cap comparable to the polar cap size determined by the open field lines. Such small Lorentz factors may be capable of explaining the gamma-ray emission from PSR 1509-58 which is observed only at energies $\lesssim 1$ MeV. We calculated the fraction of the electron's kinetic energy that is converted to gamma rays for the three gamma-ray pulsars which show thermal x-ray spectra, namely Vela, Geminga, and PSR 1055-52. Using the pulsar parameters derived by Ögelman (1995), we found that we can expect these pulsars to have between $\sim 5\%$ (Geminga) and $\sim 60\%$ (Vela) of the accelerated electron luminosity converted to gamma-ray luminosity.

## 1. Introduction

We have developed a gamma-ray pulsar model in which thermal x-ray emission from the neutron star surface is resonantly upscattered to gamma-ray energies near the pulsar polar

---

cap (Sturner, Dermer, & Michel 1995; Sturner & Dermer 1994; Dermer & Sturner 1994). The energetic gamma rays will initiate a pair cascade through $\gamma$-B pair production. This pair cascade will produce a hollow cone of emission with an opening angle $\sim$1.5 times larger than the polar cap opening angle. We have shown that such a hollow cone of emission can explain both broad single-peaked and narrow double-peaked pulse profiles if the angle between the magnetic and spin axes of the pulsar is similar to the polar cap opening angle (Sturner, Dermer, & Michel 1995; Sturner & Dermer 1994). The assumption that the young ($\leq 10^5$ year old) gamma-ray pulsars have small obliquity differs from inferences of radio pulsar astronomy (e.g. Lyne & Manchester 1988, Rankin 1993). It also precludes the observation of four gamma-ray pulse peaks. While although this pulsar geometry reduces the likelihood of detecting a given pulsar, we have found that this nearly-aligned model can account for the observed number and gamma-ray luminosities of the six known gamma-ray emitting pulsars (Dermer & Sturner 1994).

This model depends on the ability to convert the kinetic energy of accelerated electrons to gamma-ray energy through magnetic Compton scattering. Thus it is important that the six known gamma-ray pulsars (Crab, Vela, Geminga, PSR 1706-44, PSR 1055-52, and PSR 1509-58) have all been detected at soft x-ray energies (see e.g. Ögelman 1995 for a review of the observations). The emission from three of these pulsars (Vela, Geminga, and PSR 1055-52) appears to be well represented by a two-component thermal model (Ögelman 1995) which can be interpreted as thermal emission from the entire neutron star surface plus a hotter thermal polar cap (Halpern & Ruderman 1993). The temperature of the thermal polar cap is generally a few$\times 10^6$ K (Ögelman & Finley 1993; Ögelman, Finley, & Zimmermann 1993; Halpern & Ruderman 1993).

In this work we examine the conversion of electron kinetic energy to gamma-ray energy by Compton upscattering soft x-ray photons emitted from the pulsar polar cap region. In §2 we extend the calculations of Dermer (1990) to larger electron Lorentz factors ($\gamma \lesssim 10^8$) and compare our results with those of Kardeshëv, Mitrofanov, & Novikov (1984), Daugherty & Harding (1989), and Chang (1995). We include the Klein-Nishina correction to the magnetic Compton cross section, triplet pair production, and curvature radiation energy losses. In §§3 and 4 we calculate the electron Lorentz factor as a function of height above the pulsar polar cap as well as the fraction of the electron's energy that is converted to gamma rays using the particle acceleration model of Michel (1974) which was developed in detail by Fawley, Arons, & Scharlemann (1977). We examine the effects of surface magnetic field strength, polar cap temperature, and thermal polar cap size on the energy loss fraction and the electron Lorentz factor that is attained.

## 2. Electron Energy Loss Rates

The electron energy loss rate due to Compton upscattering soft photons emitted from a pulsar thermal polar cap can be expressed as

$$-\dot\gamma_{\rm Comp} = c \int_0^\infty d\epsilon \int_{\rm polar\ cap} d\Omega\ n_{\rm ph}(\epsilon,\Omega)(1-\beta\cos\Psi) \int_0^{\epsilon'_s} d\epsilon'_s \oint d\Omega'_s \left(\frac{d\sigma}{d\epsilon'_s d\Omega'_s}\right)(\epsilon_s - \epsilon) \quad (1)$$

where $n_{\rm ph}(\epsilon, \Omega)d\epsilon d\Omega$ is the density of soft photons with energy between $\epsilon$ and $\epsilon + d\epsilon$ within solid angle $d\Omega$ from direction $\vec{\Omega} = (\theta = \cos^{-1}\mu, \phi)$ with respect to the local magnetic field direction, $d\sigma/d\epsilon'_{\rm s}d\Omega'_{\rm s}$ is the differential Compton cross section in the electron rest frame, $\epsilon$ is the photon energy in units of $m_{\rm e}c^2$, and $\Psi$ is the angle between the soft photon and scattering electron momenta. We use the convention that primes denote quantities in the electron rest frame and "s" subscripts denote scattered photon quantities. Note that when we use the term thermal polar cap in this work we refer to the region of the neutron star surface from which the hotter thermal component presumably originates. The size of this region may not be the same size as the region containing the footprints of open magnetic field lines which is traditionally called the polar cap region.

Photons that have an energy in the electron rest frame equal to the local cyclotron energy experience a greatly enhanced probability of scattering because in the large magnetic fields associated with neutron stars, the Compton cross section has a resonance at the local cyclotron energy (Herold 1979), i.e. when $\epsilon' = \epsilon\gamma(1 - \beta\mu) = \epsilon_{\rm B}$ where $\gamma$ is the electron Lorentz factor, $\beta$ is the ratio of the electron velocity to the speed of light, $\epsilon_{\rm B} = B/B_{\rm cr}$, and $B_{\rm cr} = 4.414 \times 10^{13}$ G. The polarization-averaged differential magnetic Compton cross section in the Thomson regime ($\epsilon' \ll 1$) can be written as (Dermer 1990)

$$\frac{d\sigma}{d\epsilon'_{\rm s}d\Omega'_{\rm s}} = \frac{3}{16\pi}\sigma_{\rm T}\delta(\epsilon'_{\rm s} - \epsilon')\left[(1 - \mu'^2)(1 - \mu'^2_{\rm s}) + \frac{1}{2}(1 + \mu'^2)(1 + \mu'^2_{\rm s})(g_1 + 323\ \delta(\epsilon' - \epsilon_{\rm B}))\right], \quad (2)$$

where $\sigma_{\rm T}$ is the Thomson cross section, $g_1 = u^2/(u+1)^2$, and $u = \epsilon'/\epsilon_{\rm B}$. Using the approximation $(1 + \mu'^2) \cong 2$ and averaging over the angular distribution of scattered photons to simplify the second term in equation (2), we can approximate the magnetic Compton cross section as

$$\frac{d\sigma}{d\epsilon'_{\rm s}d\Omega'_{\rm s}} \approx \frac{1}{4\pi}\sigma_{\rm T}\delta(\epsilon'_{\rm s} - \epsilon')\left[\frac{3}{4}(1 - \mu'^2)(1 - \mu'^2_{\rm s}) + g_1 + 323\ \delta(\epsilon' - \epsilon_{\rm B})\right], \quad (3)$$

where the three terms in the braces can be called the angular, nonresonant, and resonant portions of the magnetic Compton cross section, respectively. We find that $|\mu'| = |(\mu - \beta)/(1 - \beta\mu)| \approx 1$ is a good approximation in view of the resonance condition given above, because the nonresonant and resonant terms in the cross section are important only when $\beta \approx 1$ ($\gamma \gg 1$) given polar cap temperatures of a few$\times 10^6$ K and magnetic field strengths $\gtrsim 10^{12}$ G.

To simplify our calculations, we will assume that the electron momenta are parallel to the magnetic field direction. This assumption is valid because of the short timescales for electrons to lose their momentum perpendicular to the magnetic field direction through synchrotron radiation in the terragauss magnetic fields associated with pulsars. Secondly, we will take the electrons to be traveling along the magnetic axis field line. These assumptions allow us to replace $\cos\Psi$ with $\mu$ in equation (1) and will simplify the $d\Omega$ integration because of the assumed azimuthal symmetry of the soft photon source with respect to the magnetic axis. Dermer (1990) has calculated the electron energy loss rate due to magnetic Compton scattering using equations (1) and (3) for the case of electrons traveling along the magnetic axis of a neutron star emitting blackbody radiation. At height $h$ above a pulsar polar cap emitting a blackbody photon distribution with temperature $T = 10^6 T_6$ K and with a surface magnetic field strength $B = 10^{12} B_{12}$ G, the energy loss rates for the three parts of the cross section are

$$-\dot{\gamma}_{\rm ang} = 46.1 \, T_6^4 (1 - \mu_c) f_{\rm ang} \quad {\rm s}^{-1}, \qquad (4)$$

$$-\dot{\gamma}_{\rm nres} = 0.0192 \left( \frac{T_6^6 \gamma^4 f_{\rm nres}}{\beta B_{12}^2(h)} \right) \quad {\rm s}^{-1}, \qquad (5)$$

$$-\dot{\gamma}_{\rm res} = 4.92 \times 10^{11} \left( \frac{T_6 B_{12}^2(h) f_{\rm res}}{\beta \gamma} \right) \quad {\rm s}^{-1}, \qquad (6)$$

where

$$f_{\rm ang} = \frac{\gamma^2 - 2}{\gamma^2 - 1} - \frac{1 - \mu_c^3}{3(1 - \mu_c)} - \frac{1 + \mu_c}{2\beta\gamma^2} + \frac{\ln[(1 - \beta\mu_c)/(1 - \beta)]}{\beta^3 \gamma^4 (1 - \mu_c)}, \qquad (7)$$

$$f_{\rm nres} = (1 - \beta\mu_c)^5 - (1 - \beta)^5 - \left\{ \frac{5[(1 - \beta\mu_c)^4 - (1 - \beta)^4]}{4\gamma^2} \right\}, \qquad (8)$$

$$f_{\rm res} = -\ln[1 - \exp(-w)]. \qquad (9)$$

Here $w = \epsilon_{\rm B}/[\gamma\theta(1 - \beta\mu_c)]$, $\theta = kT/m_e c^2$, $B_{12}(h)$ is the magnetic field strength at height $h$ in units of $10^{12}$ G, and $\mu_c = \cos\theta_c = h/\sqrt{h^2 + R_{\rm therm}^2}$ where $R_{\rm therm}$ is the radius of the thermal polar cap (see Figure 1). Equation (5) was calculated using the approximation that $g_1 = u^2$ and is thus only applicable when $\epsilon' < \epsilon_{\rm B}$. When $\epsilon' > \epsilon_{\rm B}$, $g_1 \cong 1$ and thus the nonresonant portion of the cross section is equal to the Thomson cross section.

Since the temperature of the hot thermal polar cap component is generally a few $\times 10^6$ K, we find that $\epsilon' \sim \gamma\epsilon > 1$ for typical thermal photon energies, $\sim kT$, when $\gamma \gtrsim 10^3$. For these conditions Klein-Nishina effects must be taken into account. This is one area where we improve over the previous treatments of Dermer (1990) and Chang (1995). The polarization-averaged differential Klein-Nishina cross section can be written as

$$\frac{d\sigma}{d\epsilon'_s d\Omega'_s} = \frac{3\sigma_{\rm T}}{16\pi} \left( \frac{\epsilon'_s}{\epsilon'} \right)^2 \left[ \frac{\epsilon'}{\epsilon'_s} + \frac{\epsilon'_s}{\epsilon'} - \sin^2\chi'_s \right] \delta\left( \epsilon'_s - \frac{\epsilon'}{1 + \epsilon'(1 - \cos\chi'_s)} \right), \qquad (10)$$

where $\chi'_s$ is the angle between the propagation directions of the incoming soft photon and the outgoing scattered photon in the electron rest frame. Substituting equation (10) into equation (1) and integrating we find

$$-\dot{\gamma}_{\rm KN} = \frac{3}{8} c\sigma_{\rm T} \gamma \int_0^\infty d\epsilon \int_{\rm polar\ cap} d\Omega \, n_{\rm ph}(\epsilon, \Omega)(1 - \beta\mu) \left[ \frac{\epsilon'(1 + \beta\mu') + \beta\mu'}{\epsilon'} \right] f_{\rm KN}, \qquad (11)$$

where

$$f_{\rm KN} = \frac{2(\epsilon'^2 - 4\epsilon' - 3)}{\epsilon'(1+2\epsilon')} + \frac{2\epsilon'}{3}\left[\frac{4\epsilon'^2 + 6\epsilon' + 3}{(1+2\epsilon')^3}\right] - \frac{2(1+\epsilon')(\epsilon'^2 - 2\epsilon' - 1)}{\epsilon'(1+2\epsilon')^2} - \left[\frac{\epsilon'^2 - 2\epsilon' - 3}{\epsilon'^2}\right]\ln(1+2\epsilon') - \frac{2}{\epsilon'}. \tag{12}$$

For simplicity we now take the polar cap soft photon emission to be mono-energetic with $\epsilon = \epsilon_{\rm o} = 2.7\theta$. Thus we approximate the differential photon density as

$$n_{\rm ph}^{\rm app}(\epsilon, \Omega) = \frac{\delta(\epsilon - \epsilon_{\rm o})}{\epsilon_{\rm o} m_e c^3} \int_0^\infty B_\nu \, d\nu = \frac{\sigma_{\rm B} T^3}{2.7 \pi k c} \delta(\epsilon - \epsilon_{\rm o}), \tag{13}$$

where $B_\nu$ is the Planck function and $\sigma_{\rm B}$ is the Stefan-Boltzmann constant. Inserting equation (13) into equation (11) and integrating, we find that

$$\begin{aligned} -\dot{\gamma}_{\rm KN} &= \frac{3\sigma_{\rm T} \sigma_{\rm B} m_e^2 c^4 T}{4\beta\gamma^2 (2.7)^3 k^3} \int_{\epsilon'_-}^{\epsilon'_+} d\epsilon' \left(\epsilon_{\rm o} + \frac{\epsilon_{\rm o}}{\epsilon'} - \gamma\right) f_{\rm KN} \\ &= 3.7 \times 10^{11} \left(\frac{T_6}{\beta\gamma^2}\right) \int_{\epsilon'_-}^{\epsilon'_+} d\epsilon' \left(\epsilon_{\rm o} + \frac{\epsilon_{\rm o}}{\epsilon'} - \gamma\right) f_{\rm KN} \quad {\rm s}^{-1}, \end{aligned} \tag{14}$$

where $\epsilon'_- = \epsilon_{\rm o}/\gamma(1+\beta)$, $\epsilon'_+ = \epsilon_{\rm o}/\gamma(1+\beta\mu'_{\rm c})$, and $\mu'_{\rm c} = (\mu_{\rm c} - \beta)/(1 - \beta\mu_{\rm c})$.

When the Lorentz factors of the electrons exceed $\sim 10^3$, other processes may also be important. One such process is triplet pair production ($e\gamma \to ee^+e^-$) in which a photon interacts with the coulomb field of the electron, producing a positron-electron pair in addition to the original electron (see e.g. Mastichiadis 1991; Mastichiadis, Marscher, & Brecher 1986). The energy loss rate for triplet pair production in an isotropic radiation field of mono-energetic soft photons has been calculated by Dermer & Schlickeiser (1991) under the assumptions that $\epsilon \ll 1 \ll \gamma$, which is generally true for conditions we investigate in this work, and using the analytic approximation that the energy of the formed electron or positron $\approx \gamma/\sqrt{2\epsilon'}$. They found that their analytic solution was accurate to better than a factor of 3 for $10 \lesssim \gamma\epsilon \lesssim 10^7$ when compared with the exact numerical solution of Mastichiadis (1991). Their solution tended to slightly overestimate the electron energy loss rates for small and large values of $\gamma\epsilon$.

Following their analysis, the energy loss rate via triplet pair production in the non-isotropic photon field along the magnetic axis near a pulsar polar cap, assuming monoenergetic soft photons with $\epsilon = \epsilon_{\rm o} = 2.7\theta$, can be written as

$$-\dot{\gamma}_{\rm tpp} = c \int_0^\infty d\epsilon \int_0^{2\pi} d\phi \int_{\mu_{\rm c}}^{\mu_{\rm thr}} d\mu \, n_{\rm ph}(\epsilon, \Omega)(1-\beta\mu)\sigma_{\rm tpp}\Delta\gamma, \tag{15}$$

where $n_{\rm ph}(\epsilon, \Omega)$ is the differential, monoenergetic photon density given by equation (13), $\sigma_{\rm tpp} \approx \alpha_{\rm f} \sigma_{\rm T}$ is the triplet pair production cross section, $\alpha_{\rm f}$ is the fine structure constant, $\Delta\gamma \approx 2\gamma/\sqrt{2\epsilon'}$ is the energy lost by the electron per interaction, and $\mu_{\rm thr} = (1 - 4/\epsilon_{\rm o}\gamma)/\beta$ is the cosine of the angle at which $\epsilon' = 4$, the threshold energy for the process to occur. Integrating equation (15), we find the energy loss rate for triplet pair production is

$$
\begin{aligned}
-\dot{\gamma}_{\rm tpp} &\approx \left(\frac{0.0051\sigma_{\rm T}\sigma_{\rm B}T^3}{k\beta\gamma\epsilon_{\rm o}^2}\right)\left\{[\epsilon_{\rm o}\gamma(1-\beta\mu_{\rm c})]^{3/2} - 8\right\}, \\
&\approx 6.7 \times 10^9 \left[\frac{T_6^3}{\beta\gamma}\right]\left\{\left[4.5 \times 10^{-4}T_6\gamma(1-\beta\mu_{\rm c})\right]^{3/2} - 8\right\} \ {\rm s}^{-1}.
\end{aligned}
\tag{16}
$$

The threshold electron Lorentz factor for this process is found by setting the term in braces to zero, i.e. setting $\mu_{\rm c} = \mu_{\rm thr}$, yielding

$$
\gamma_{\rm tpp,thr}\left(1 - \beta_{\rm tpp,thr}\,\mu_{\rm c}\right) = \frac{4}{\epsilon_{\rm o}} \approx \frac{8800}{T_6}. \tag{17}
$$

When the electron Lorentz factors are even larger, $>{\rm few}\times 10^6$, curvature radiation losses may become important. The energy loss rate due to curvature radiation emission is the same as for synchrotron radiation except the radius of curvature of the magnetic field line is used instead of the gyro radius. Thus the energy loss rate is

$$
-\dot{\gamma}_{\rm curv} = \frac{2e^2\beta^4\gamma^4}{3m_{\rm e}c\rho_{\rm B}^2} \approx 5.6 \times 10^{-3}\left(\frac{\gamma^4}{\rho_{\rm B}^2}\right) \ {\rm s}^{-1}, \tag{18}
$$

where $\rho_{\rm B}$ is the radius of curvature of the magnetic field line. As stated above, the energy loss rates for Compton scattering and triplet pair production have been calculated along the magnetic axis field line. This was done because the azimuthal symmetry greatly simplifies the calculations. The magnetic axis field line in a dipole magnetic field is not expected to have any curvature if the magnetic and spin axes of the pulsar are aligned. If they are not aligned however, some curvature of the polar field line may be present. We also expect that the magnetic Compton and triplet pair production solutions should not vary significantly when the electron is slightly off the magnetic axis. The lowest order correction to the magnetic Compton and triplet pair production energy loss rates for an electron slightly off the magnetic axis at magnetic colatitude $\theta_{\rm m}$ would be the inclusion of a factor of $\cos\theta_{\rm m}$ to account for the decrease in the soft photon density due to the decreased solid angle subtended by the polar cap. For $\theta_{\rm m} < 10°$, this results in only a $\lesssim 2\%$ reduction in the photon density and thus also in the magnetic Compton and triplet pair production energy loss rates. The value of $\rho_{\rm B}$ goes from infinity at $\theta_{\rm m} = 0$ for a pure dipole magnetic field to $\rho_{\rm B} \sim 10^7$ cm for $5° \lesssim \theta_{\rm m} \lesssim 10°$ (see e.g. Arons 1983). Thus in this work we will use the energy loss rates derived above and use a finite magnetic field line radius of curvature.

We can use these electron energy loss rates to calculate an energy loss scale length, $\lambda_{-,{\rm i}} \cong c\beta\gamma/\dot{\gamma}_{\rm i}$, over which the electron will lose $\sim 1/e$ of its energy. In Figures 2a and 2b we show the energy loss rates and energy loss scale lengths at the neutron star surface (i.e. $h = 0$

and $\mu_c = 0$) for the various processes as a function of electron Lorentz factor for $B_{12} = 3.5$ and $T_6 = 3.5$. The results shown in Figure 2a are very similar to those presented by Chang (1995) for the magnetic Compton energy loss rates when $\gamma \lesssim 10^3$. They also agree qualitatively with Figure 1 of Kardeshëv, Mitrofanov, & Novikov (1984) in which the authors sketch the radiative braking forces due to these processes, except we find that there is no range of electron Lorentz factors for which triplet pair production losses dominate given these parameter values. This difference occurs because they calculated that the Klein-Nishina and triplet pair production braking forces are equal when $\gamma\theta \sim 1/\alpha_f$ by comparing the cross sections for the two processes. They did not take into account that the energy transfer per interaction is greater for Compton scattering in the Klein-Nishina regime, $\Delta\gamma \sim \gamma$, than for triplet pair production, $\Delta\gamma \sim \sqrt{2\gamma/\theta}$, when $\gamma > 2/\theta$. Dermer & Schlickeiser (1991) found that the energy loss rates for these two processes are equal when $\gamma\theta \sim 10^5$.

Note that resonant Compton scattering dominates when the Lorentz factors are between 10 and $10^4$. The local minimum of $\lambda_{-,\text{tot}}$ due to the resonance in the magnetic Compton cross section occurs at an electron Lorentz factor

$$\gamma_{\text{res}} \cong \frac{\epsilon_B}{\epsilon_o} \approx 50 \left(\frac{B_{12}}{T_6}\right). \qquad (19)$$

We find that curvature radiation losses dominate only when the electron Lorentz factors are $\gtrsim$ few$\times 10^6$ for a typical value of $\rho_B \sim 10^7$ cm. Note also the significant effect of the Klein-Nishina correction to the Compton cross section on the total energy loss scale length. It is greatly increased over the Thomson value when the Lorentz factors are $\gtrsim 10^3$. We also find that for typical pulsar parameters, triplet pair production appears only to be important for electrons with large Lorentz factors ($\gtrsim 10^8$) when either the radius of curvature of the magnetic field line is very large ($\gg 10^7$ cm) and/or the polar cap temperature is $\gg 3.5 \times 10^6$ K.

In Figures 3a and 3b we illustrate how the energy loss scale length at the neutron star surface varies with thermal polar cap temperature and surface magnetic field strength. The magnetic field strengths in Figures 3a and 3b are 3.5 and $15.8 \times 10^{12}$ G, respectively, and the temperature is varied from 2.0 to $7.0 \times 10^6$ K. Note that as the thermal polar cap temperature is increased, the number of thermal photons at all energies increases, and thus the energy loss scale length decreases for all Lorentz factors where Compton scattering losses dominate. When the polar cap temperature is raised at a fixed magnetic field strength, the location of the local minimum of $\lambda_{-,\text{tot}}$, due to the cyclotron resonance, occurs at lower Lorentz factors as is given by equation (19). That is because when the temperature is increased, the photon energy at maximum flux (and thus minimum $\lambda_-$) increases so that the Lorentz factor of the electrons must decrease in order for these higher energy photons to be in resonance with electrons. When the surface magnetic field strength is increased at constant temperature, the local minimum of the energy loss scale length moves to larger electron Lorentz factors since the photons at the peak of the thermal spectrum must have a larger doppler boost to be in resonance with electrons in the stronger magnetic field. This behavior was also noted by Daugherty & Harding (1989) when they calculated the electron scattering mean free path. In the next section we show that the location of the local minimum of $\lambda_-$ strongly affects the final energy of the electrons.

### 3. Electron Acceleration Model

In our gamma-ray pulsar model (Sturner, Dermer, & Michel 1995; Sturner & Dermer 1994; Dermer & Sturner 1994) we have used a particle acceleration model discussed in Michel (1974) and Fawley, Arons, & Scharleman (1977) and later reviewed by Michel (1982, 1991). In this model, electrons (or positrons) are accelerated away from the pulsar polar cap along open field lines by an electric field parallel to the magnetic field. This electric field originates because slight deviations from the Goldreich-Julian density develop as charges leave the polar cap area. The accelerating electric field in this model is shorted out at a height, $h_\mathrm{o}$, above the polar cap due to readjustments of the charge density in the corotating portion of the magnetosphere. This height is $\approx r_\mathrm{pc}$, the polar cap radius as defined by the footprints of the open magnetic field lines. For an aligned rotator

$$h_\mathrm{o} \cong r_\mathrm{pc} = (2\pi a^3/cP)^{1/2} = 1.45 \times 10^4 a_6^{3/2} P^{-1/2} \quad \mathrm{cm}, \tag{20}$$

where $a_6$ is the neutron star radius in units of $10^6$ cm and $P$ is the pulsar period. Michel (1982, 1991) calculates an acceleration scale length

$$\lambda_+ = \left(\frac{m_\mathrm{e} c^3 P}{8\pi e B(h)}\right)^{1/2} \approx 1.43 \left(\frac{P}{B_{12}}\right)^{1/2}, \tag{21}$$

since $h_\mathrm{o} \ll a$. The electron Lorentz factor will increase by 1 while traversing a distance $\approx \lambda_+$ in the electric field. Thus the maximum electron Lorentz factor with no losses is $\gamma_\mathrm{o} = h_\mathrm{o}/\lambda_+ \approx 10^4 \, a_6^{3/2} P^{-1} B_{12}^{1/2}$. For the six known gamma-ray pulsars this maximum Lorentz factor varies from $5.4 \times 10^4$ for PSR 1055-51 to $6.0 \times 10^5$ for the Crab pulsar (see Dermer & Sturner 1994).

In Figures 3a and 3b we also show the electron Lorentz factor as a function of distance traveled, $d$, for a pulsar period of 0.150 seconds assuming no losses. Note that in some cases $\lambda_{-,\mathrm{tot}}(\gamma) \sim d(\gamma) = \gamma \lambda_+$. When this occurs, the distance an electron must travel in the accelerating electric field to gain $\gamma m_\mathrm{e} c^2$ of energy is roughly equal to the distance needed to lose that much energy. Thus energy gains balance energy losses and an equilibrium Lorentz factor, $\bar{\gamma}$, is reached. Looking at Figures 3a and 3b we would expect there to be regions above the pulsar surface where the electron Lorentz factors remain roughly constant with values between $10^2$ and $10^3$ when $T \gtrsim 7.0 \times 10^6$ K and $B = 3.5 \times 10^{12}$ G as well as when $T \gtrsim 3.5 \times 10^6$ K and $B = 1.58 \times 10^{13}$ G. Note that for larger polar magnetic field strengths, electron acceleration can be balanced by losses at lower polar cap temperatures. We discuss this in more detail in §4.

## 4. Conversion of Kinetic Energy to Gamma Rays

Here we model the acceleration of electrons away from the neutron star surface along the magnetic axis. An electron is accelerated away from the neutron star surface by the electric field discussed in §3. In each distance step of size $dh$, the energy gained, $dh/\lambda_+$, is added to that lost, $\gamma dh/\lambda_-$, where $\lambda_-$ is derived from equations (4)-(6) when $w > 2$ and from equations (6), (14), (16), and (18) when $w < 2$, except when $\gamma < \gamma_{\rm tpp,thr}$, in which case triplet pair production is ignored. We have introduced a radius of curvature for the magnetic field line of $10^7$ cm. We would like to point out that, as is evident from Figures 2 and 3, curvature radiation emission is an unimportant energy loss mechanism when $\gamma < 10^6$ even if $\rho_{\rm B}$ is only $10^6$ cm. Thus our choice of $\rho_{\rm B}$ has little effect on the results presented in this section given that the accelerating electric field is shut off at a height equal to $h_{\rm o}$, thus limiting $\gamma_{\rm o}$ to values $< 10^6$. When $h > h_{\rm o}$, the electron is allowed to coast out to a distance of $10^6$ cm. We find that this height is sufficient to characterize the energy loss because of the small size of the thermal polar caps and the relatively low temperatures associated with the rest of the neutron star surface, generally $\lesssim 10^6$ K (Ögelman 1994). Input parameters in this model are the thermal polar cap size, its temperature, the surface magnetic field strength, the height of the accelerating region, and the pulsar period.

In Figures 4a-c we demonstrate how $\gamma(h)$ varies with polar cap temperature, thermal polar cap size, and polar magnetic field strength assuming a pulsar period of 0.150 seconds. In Figure 4a we see that the acceleration rate of an electron is decreased only slightly by losses from resonant Compton scattering when the surface magnetic field strength is $3.5 \times 10^{12}$ G, the polar cap temperature is $\leq 3.5 \times 10^6$ K, and the thermal polar cap radius is $10^5$ cm. For these parameters, the maximum electron Lorentz factor, $\gamma_{\rm max}$, is very close to $\gamma_{\rm o}$. In Figure 4b we show $\gamma(h)$ again, using the same polar cap temperature and size as for Figure 4a but for a surface magnetic field strength of $1.58 \times 10^{13}$ G. For this magnetic field strength, we see that acceleration can be halted over a range heights above the surface ($10^2 - 10^4$ cm in this case). This effect is due to the resonant interaction of the electron with the thermal photons and was noted by Chang (1995) as well as Kardeshëv, Mitrofanov, and Novikov (1984). The electron Lorentz factor in this region, $\bar{\gamma}$, can be estimated by setting the energy loss scale length due to resonant Compton scattering as derived from equation (6) equal to $d(\gamma) = \gamma \lambda_+$ and solving the resulting transcedental function

$$\bar{\gamma} = 23.5 \, T_6 B_{12}^{3/2} P^{1/2} \ln\left[\frac{1}{1 - \exp(-w)}\right], \tag{22}$$

where $w = 134.4 B_{12}/\bar{\gamma} T_6 (1 - \mu_{\rm c})$. For the cases of $T_6 = 3.0$ and 3.5 in Figure 4b, the solutions to this equation are $\bar{\gamma} \sim 560$ and $\sim 320$, respectively, for $h = 10^3$ cm, in close agreement with the model results.

Note in Figure 4b that significant acceleration of the electron resumes when $h \gtrsim 10^4$ cm. This occurs because the energy loss scale length is constantly increasing with increasing $h$ due to the of the shrinking solid angle subtended by the thermal polar cap. Eventually, $\lambda_{-,tot} < d(\gamma)$ for all $\gamma < \gamma_o$ and acceleration of the electron resumes until $h = h_o$. We used the same parameter values to generate Figure 4c as Figure 4b except we have increased the thermal polar cap size to $3.0 \times 10^5$ cm from $10^5$ cm. The results in Figure 4c are very similar to those presented in Figure 4b except the resumption of acceleration occurs at larger values of $h$, as expected.

In Table 1 we give the fraction of the energy given to the electron by the electric field that is converted to gamma rays for the various conditions used in Figures 4a-c. This number is given by $(\gamma_o - \gamma_{\min})/\gamma_o$ where $\gamma_{\min}$ is the final electron Lorentz factor at $h = 10^6$ cm. The fraction of the electron's energy that is converted to gamma rays varies from 7.5% to ~100%. We found that these results change by <1% if the electrons are allowed to coast out to a height $h = 2.5 \times 10^6$ cm instead of $10^6$ cm. In Figures 5a-c we show how the temperature that is required to have 10%, 25%, and 99% of the energy given to the electron by the electric field converted to photons varies with surface polar magnetic field strength and thermal polar cap size including the case when the entire neutron star surface radiates. The temperature that is necessary to have a fixed fraction of the electron energy converted to photons decreases as the surface polar magnetic field strength is increased. This effect is particularly noticeable when 99% of the kinetic energy is lost. When this much energy is lost, an electron will have a $\gamma$ vs. $h$ curve similar to those shown for the case of $T_6 = 3.5$ in Figure 4b or $T_6 = 3.0$ and 3.5 in Figure 4c in which $\gamma_{\max} \ll \gamma_o$.

From these results we can predict that, according to the polar cap gamma-ray production models of Dermer & Sturner (1994) and Sturner & Dermer (1994), pulsars with large magnetic fields ( $\gtrsim 1.5 \times 10^{13}$) will not have gamma-ray emission above ~100 MeV if their polar cap temperatures are $\gtrsim 3.5 \times 10^6$ K because the Lorentz factors of their electrons will tend to be limited to values $< 10^3$. In such a model, the gamma-ray spectrum is a result of a pair cascade induced by Compton upscattered photons with energies $\gtrsim 1$ GeV. The highest energy photons that can result from resonant Compton scattering have an energy of $\approx 116 B_{13} \gamma_3$ MeV where $\gamma_3$ is the electron Lorentz factor in units of $10^3$. Thus a high magnetic field pulsar with a temperature $\gtrsim 3.5 \times 10^6$ K will not produce a Compton induced pair cascade and the gamma-ray spectrum would consist of only the very flat Compton scattered spectrum (see Sturner, Dermer, & Michel 1995; Sturner & Dermer 1995). Curvature radiation will not contribute to the gamma-ray spectrum in this case since the typical curvature radiation photon only has an energy of $3 \times 10^{-3} \gamma_3^3 \rho_{B,7}$ eV where $\rho_{B,7}$ is the radius of curvature of the magnetic field line in units of $10^7$ cm. By contrast, effects of hot polar caps are not important for polar cap models in which the acceleration zone extends to larger heights above the surface or the acceleration rate is much larger. In these cases the electrons will gain energy until their acceleration is halted by curvature radiation losses when $\gamma \gtrsim 10^6$ (e.g. Daugherty & Harding 1982, 1994).

As we have previously stated, all six known gamma-ray emitting pulsars have also been detected at x-ray energies. Three of these, Vela, Geminga, and PSR 1055-52, have spectra that can be well fit by a two component thermal model. These two components can be interpretted as emission from the small pulsar polar cap region at one temperature, $T_H$, and emission from much of the rest of the neutron star surface at a lower temperature, $T_S$. In Table 2 we list the temperatures and luminosities for the hotter polar cap component of these three pulsars derived from *ROSAT* observations and listed in Ögelman (1995). These values are very similar to those given in Ögelman & Finley (1993) and Halpern & Ruderman (1993) for PSR 1055-52 and Geminga, respectively. We also list the distances used to derive these luminosities. A thermal polar cap radius, $R_{\text{therm}} = 10^5 \, R_{\text{therm},5}$ cm, can be derived from this information assuming that the polar cap can be approximated as a flat disk using the equation

$$R_{\text{therm}} = \sqrt{\frac{L_{\text{H}}}{\pi \sigma_{\text{B}} T_{\text{H}}^4}} = 7.5 \times 10^5 \sqrt{\frac{L_{\text{H},32}}{T_6^4}} \quad \text{cm}, \tag{23}$$

where $L_{\text{H},32}$ is the luminosity of the polar cap component in units of $10^{32}$ ergs/s. We give the results of this calculation in Table 2 also. In Table 3 we list the luminosities and temperatures of the softer thermal component of these three pulsars as well as the inferred neutron star radii.

Using these parameter values, we have calculated the percentage of the electron's energy that is converted to gamma rays by magnetic Compton scattering. We find that a significant portion of the electron's kinetic energy can be converted to gamma rays by scattering the thermal polar cap emission. The percentage of the electron's kinetic energy that is converted to gamma-rays is 5% for Geminga, 20.5% for PSR 1055-52, and 60% for Vela parameters. We next consider what percentage of the electron's kinetic energy would be converted to gamma rays if the entire neutron star were at $T_{\text{S}}$, i.e. with no hot thermal polar cap component. This percentage ranges from ∼1.4% for Geminga to ∼3.2% for Vela. Thus we see that the major source of electron energy loss for these pulsars, given our particle acceleration model, is magnetic Compton scattering of the hot thermal polar cap component with Comptonization of the emission from the rest of the neutron star surface contributing at the ∼5% - 25% level depending on the size and temperature of the polar cap as well as the temperature of the rest of the neutron star surface.

## 5. Summary

We have derived the energy loss rates for a relativistic electron travelling along the symmetry axis of a thermal radiation field. The processes we investigated were magnetic Compton scattering including the Klein-Nishina correction to the Thomson cross section, triplet pair production, and curvature radiation emission. We have found that when the electron's Lorentz factor is $\lesssim$ few$\times 10^6$, magnetic Compton scattering losses dominate. At Lorentz factors $\gtrsim$ few$\times 10^6$ curvature radiation losses dominate assuming typical field line radii of curvature of $\sim 10^7$ cm. Triplet pair production losses were generally found to be unimportant for gamma-ray pulsar parameters.

We used these energy loss rates to calculate the electron Lorentz factor as a function of height above a neutron star polar cap using the electron acceleration model given in Michel (1974). We found that for a given thermal polar cap temperature and size, an electron would have more of its kinetic energy converted to gamma rays through Compton scattering when the pulsar's surface magnetic field strength was larger. In fact, we found that for a magnetic field strength similar to PSR 1509-58 ($1.58 \times 10^{13}$ G), an electron would have practically all of its kinetic energy converted to gamma-rays and the electron would be limited to a Lorentz factor between 100 and 1000 if the polar cap temperature was $\gtrsim 3.0 \times 10^6$ K, confirming the result of Chang (1995). This may lead to an explanation of why this pulsar has a gamma-ray spectrum that is qualitatively different from the other gamma-ray pulsars, in that it has only been detected at gamma-ray energies $\lesssim 1$ MeV while the others have been seen to GeV energies.

We calculated the fraction of the electron's kinetic energy that is converted to gamma rays for the three gamma-ray pulsars which show thermal x-ray spectra, namely Vela, Geminga, and PSR 1055-52. Using the pulsar parameters derived by Ögelman (1995), we found that we can expect these pulsars to have between ∼5% (Geminga) and ∼60% (Vela) of the accelerated electron luminosity converted to gamma-ray luminosity.

We would like to thank C. D. Dermer, A. K. Harding, and the anonymous referees for useful discussions. This work was supported by NASA grant 93-085.

Table 1: Percentage of Energy Lost for Parameters Used in Figures 4a-c

| Period (s) | $B_{12}$ | $R_{\text{therm},5}$ | $T_6$ | $\gamma_{\max}(10^5)$ | % Energy Lost |
|---|---|---|---|---|---|
| 0.150 | 3.5 | 1.0 | 2.0 | 1.22 | 7.5 |
|  |  |  | 2.5 | 1.20 | 11.4 |
|  |  |  | 3.0 | 1.17 | 16.3 |
|  |  |  | 3.5 | 1.14 | 22.4 |
| 0.150 | 15.8 | 1.0 | 2.0 | 2.45 | 17.7 |
|  |  |  | 2.5 | 2.31 | 27.2 |
|  |  |  | 3.0 | 1.30 | 80.5 |
|  |  |  | 3.5 | 0.04 | 99.97 |
| 0.150 | 15.8 | 3.0 | 2.0 | 2.44 | 25.6 |
|  |  |  | 2.5 | 2.29 | 39.7 |
|  |  |  | 3.0 | 0.45 | 99.98 |
|  |  |  | 3.5 | 0.004 | 99.99 |

Table 2: Thermal Polar Cap Parameters

| PSR B | Period (s) | $B_{12}$ | d (kpc) | $T_{\rm H}$ ($10^6$ K) | $L_{\rm H,32}$ | $R_{\rm therm,5}$ | % Energy Lost |
|---|---|---|---|---|---|---|---|
| Vela | 0.089 | 3.5 | 0.50 | 15.9 | 10 | 0.1 | 59.6 |
| Geminga | 0.237 | 1.7 | 0.25 | 3.2 | 0.02 | 0.1 | 5.0 |
| 1055-52 | 0.197 | 1.1 | 1.00 | 2.5 | 2.0 | 1.7 | 20.5 |

Table 3: Soft Thermal Component Parameters

| PSR B | d (kpc) | $T_S$ ($10^6$ K) | $L_{S,32}$ | $a_6$ | % Energy Lost |
|---|---|---|---|---|---|
| Vela | 0.50 | 1.26 | 6.3 | 0.6 | 3.2 |
| Geminga | 0.25 | 0.50 | 0.5 | 1.0 | 1.4 |
| 1055-52 | 1.00 | 0.63 | 15.8 | 3.0 | 3.1 |

## Figure Captions

Fig. 1: Polar cap geometry.

Fig. 2: The energy loss rate (a) and energy loss length scale (b) at the neutron star surface as a function of electron energy for various processes when $B_{12} = 3.5$ G, $T_6 = 3.5$ K, and $\rho_B = 10^7$ cm. Note how resonant Compton scattering dominates at electron Lorentz factors between 10 and $10^4$. Curvature radiation losses dominate when $\gamma > 10^6$.

Fig. 3: The energy loss length scale at the neutron star surface as a function of electron energy when $B_{12} = 3.5$ (a) and 15.8 (b), $\rho_B = 10^7$, and $T_6 =$ 2.0 (short-dashed), 3.5 (long-dashed), and 7.0 (solid). Also shown is the distance, $d$, an electron must travel to reach a given Lorentz factor according to the acceleration model of Michel (1974) for P = 0.150 seconds. Note that the local minimum due to the resonance in the magnetic Compton cross section occurs at larger Lorentz factors for larger magnetic field strengths.

Fig. 4: The Lorentz factor of an electron accelerated away from the pulsar polar cap as a function of height. Here we chose $P = 0.150$ seconds, $\rho_B = 10^7$, $B_{12} = 3.5$ (a) and 15.8 (b, c), and $R_{\text{therm},5} = 1.0$ (a, b) and 3.0 (c). The polar cap temperature was varied from 2.0 to $3.5 \times 10^6$ K. The bend near $\gamma = 100$ in (a) is due to the resonance in the magnetic Compton cross section. Note how the resonance temporarily halts acceleration when $T_6 \geq 3.0$ in (b) and (c), and how the height at which acceleration resumes after it has been halted is larger when the thermal polar cap is larger.

Fig. 5: The polar cap temperature as a function of surface magnetic field strength and thermal polar cap size that is necessary to have 10%, 25%, and 99% of the electron's kinetic energy converted to gamma-rays. Here we have taken the pulsar period to be 0.150 seconds. Note that as the magnetic field strength is increased, the temperature that is necessary to convert a given fraction of the electron's kinetic energy into gamma rays is lower.

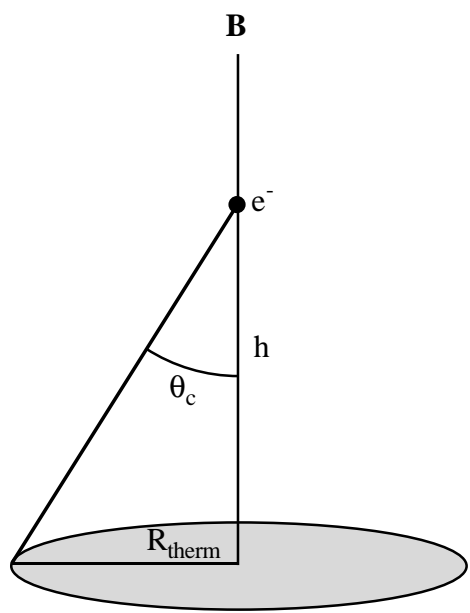

Figure 1

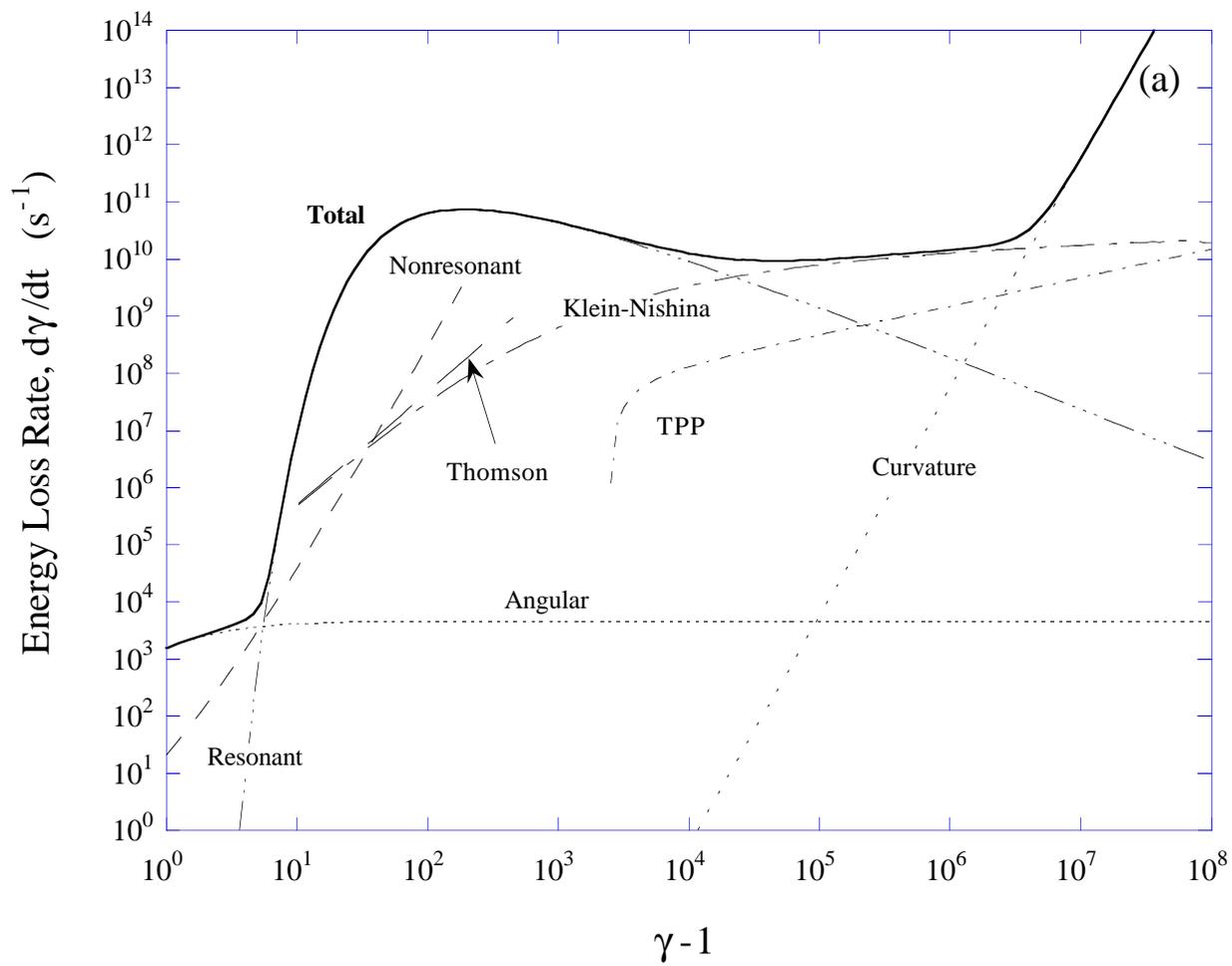

Figure 2a

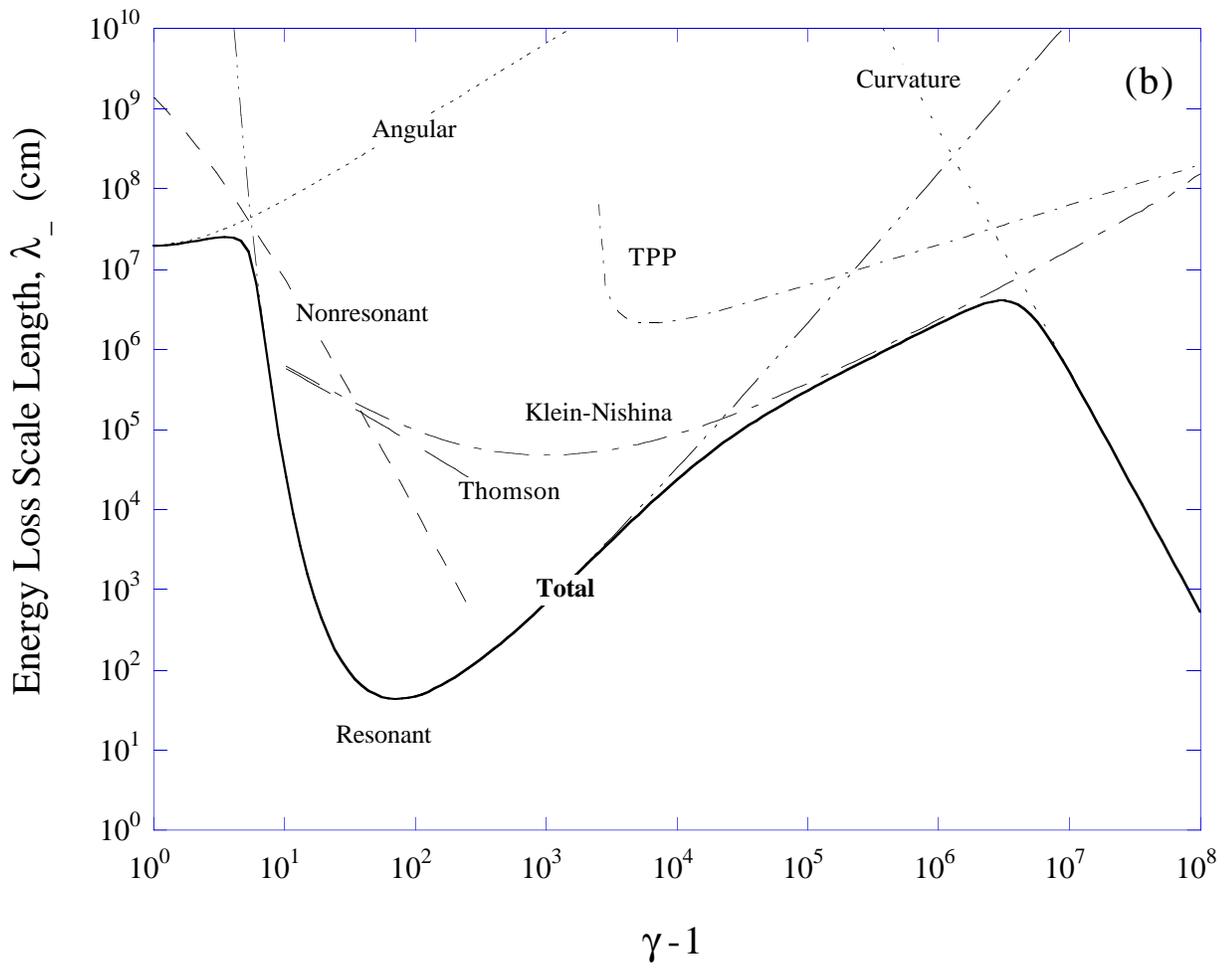

Figure 2b

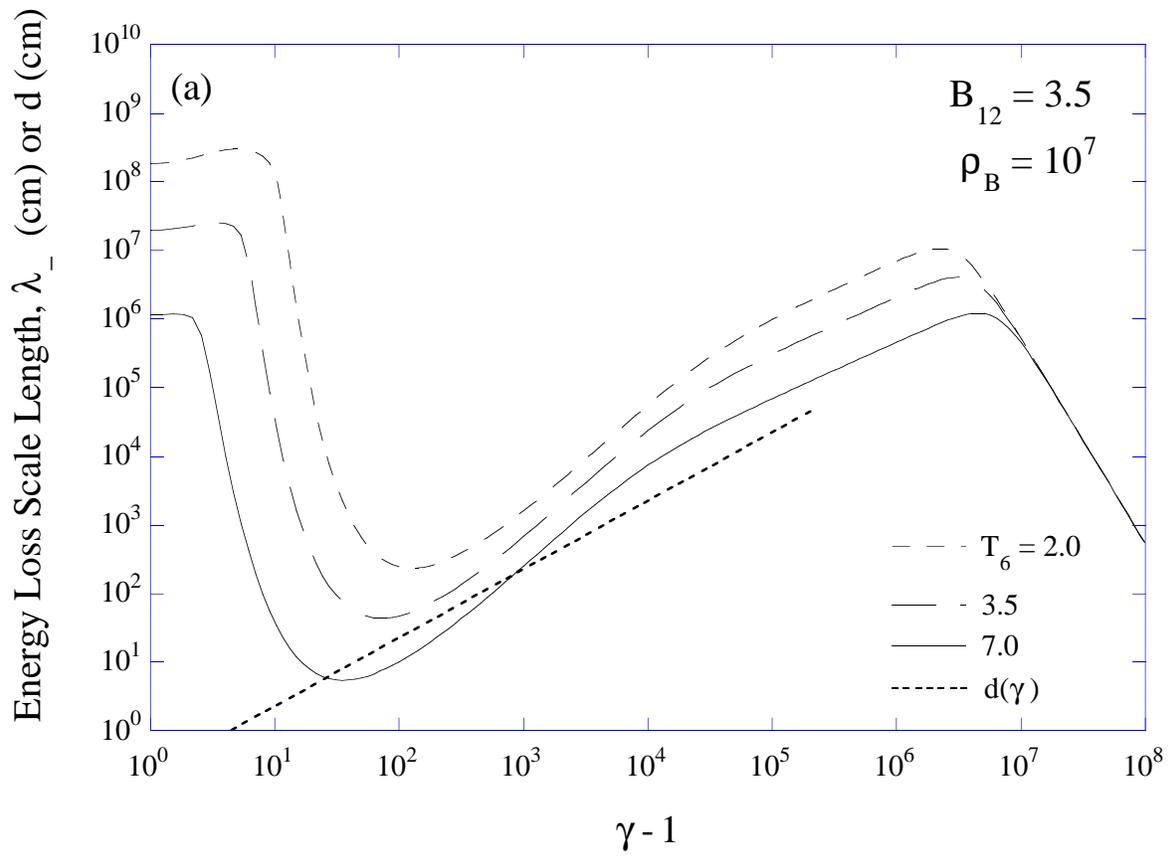

Figure 3a

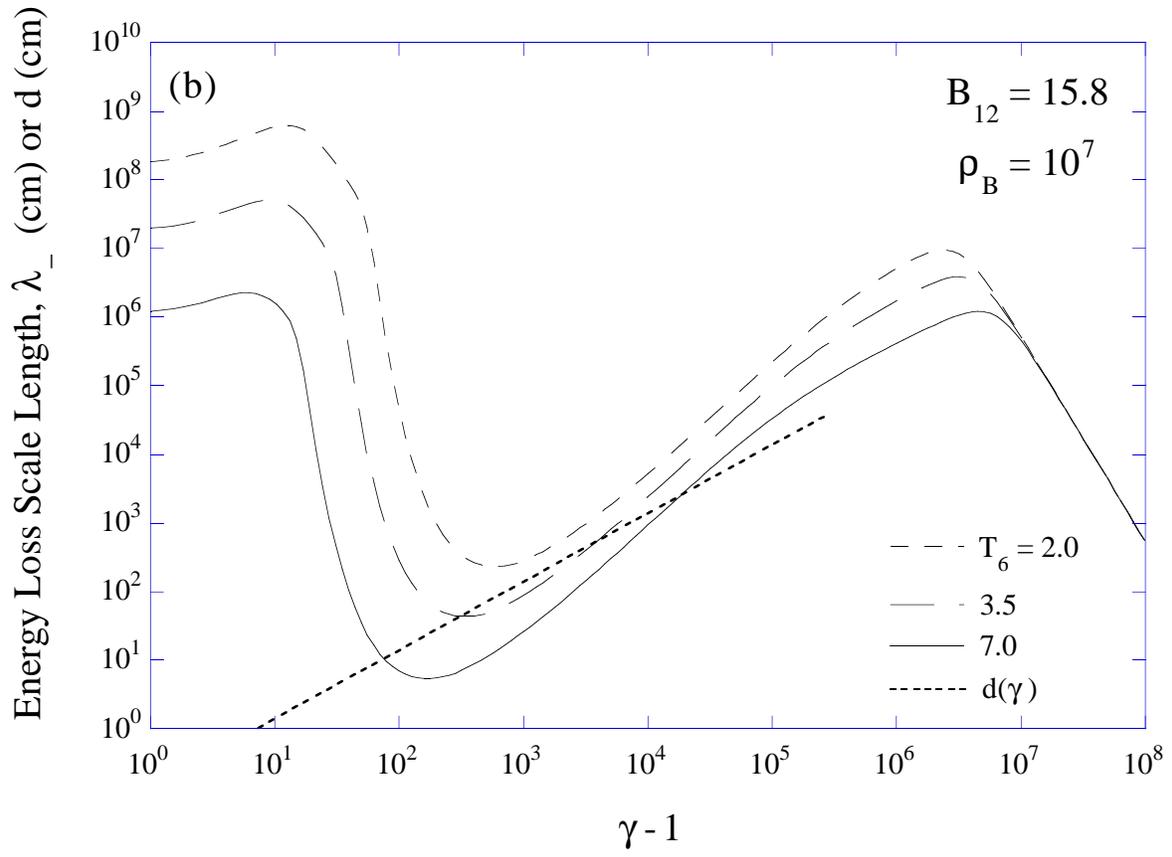

Figure 3b

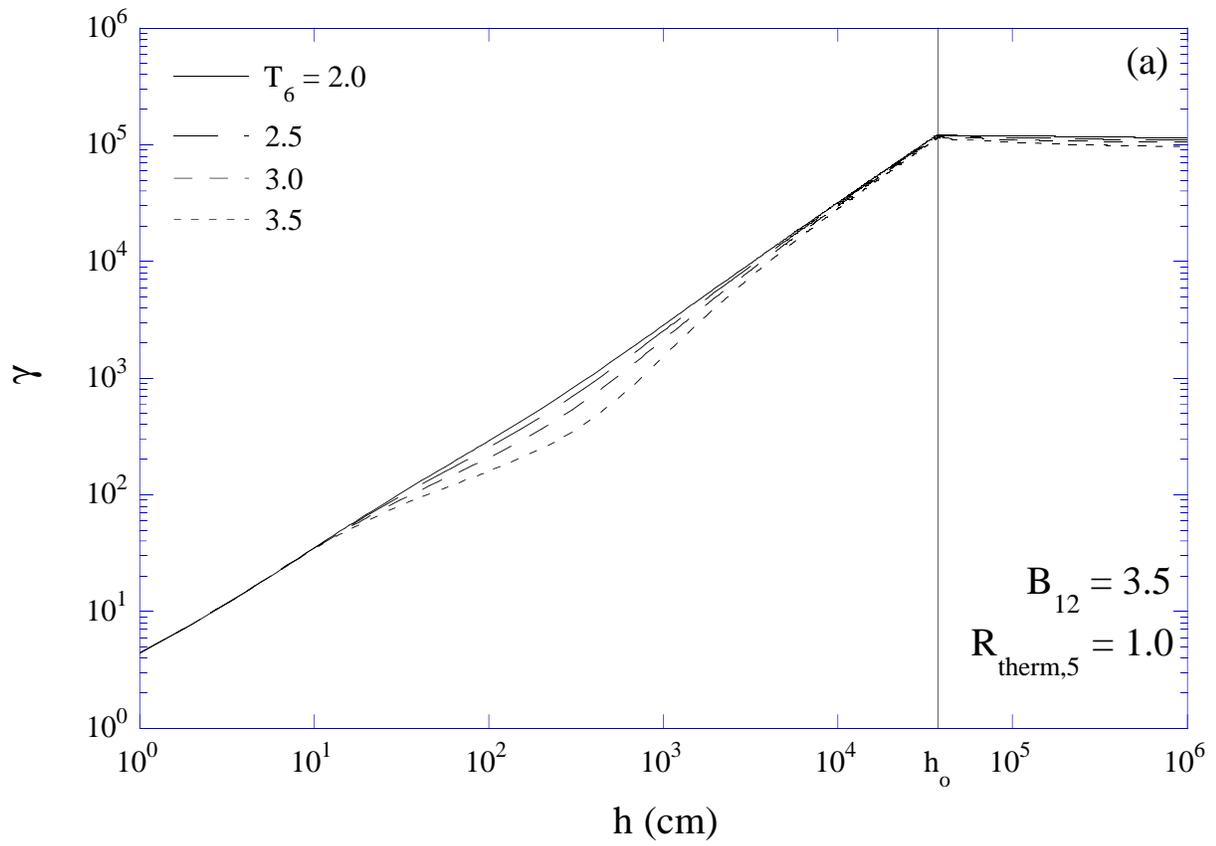

Figure 4a

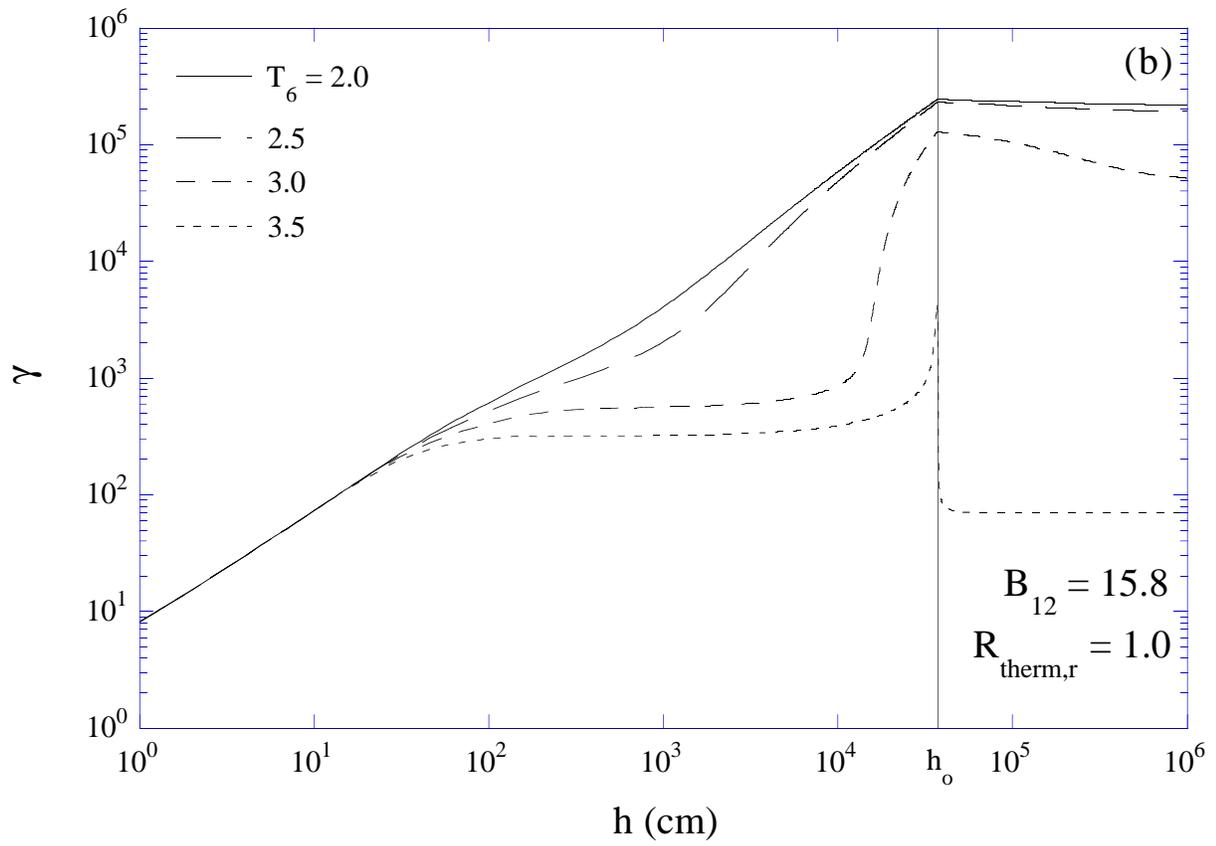

Figure 4b

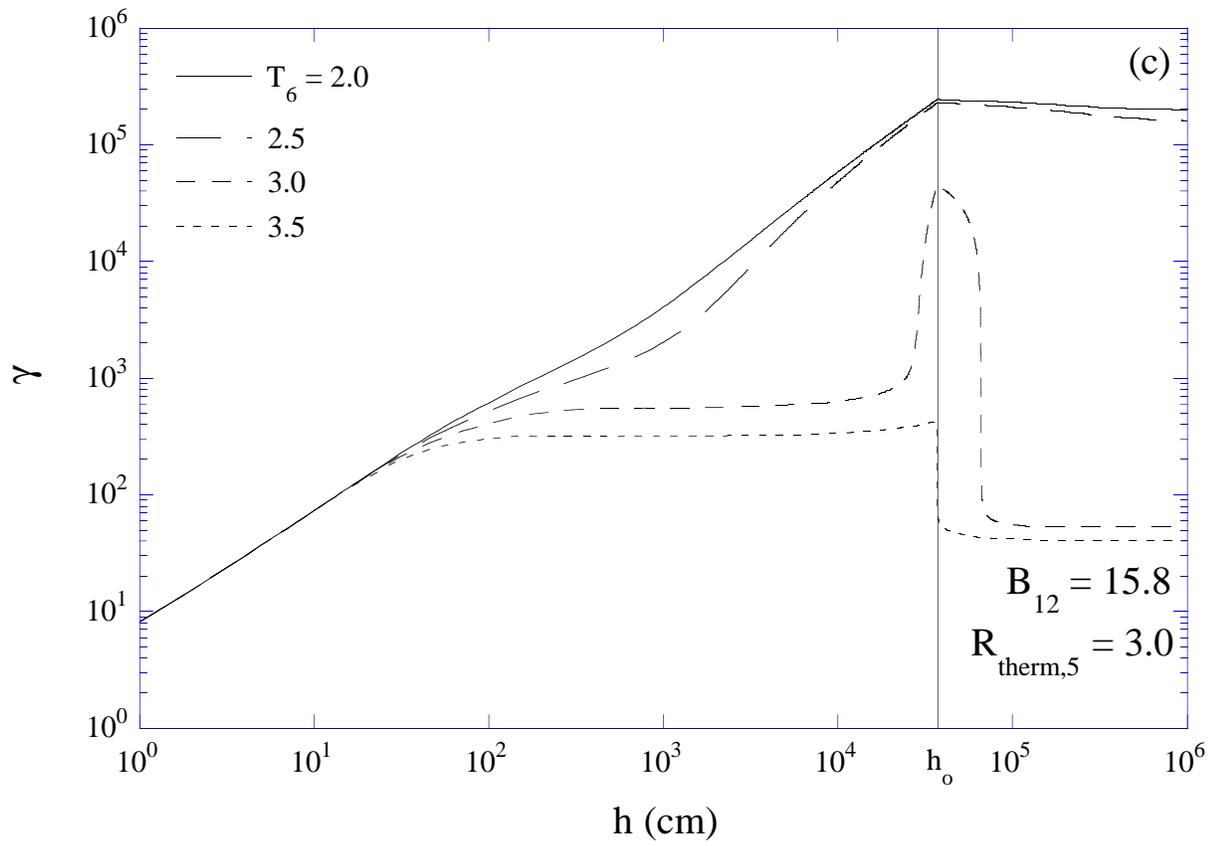

Figure 4c

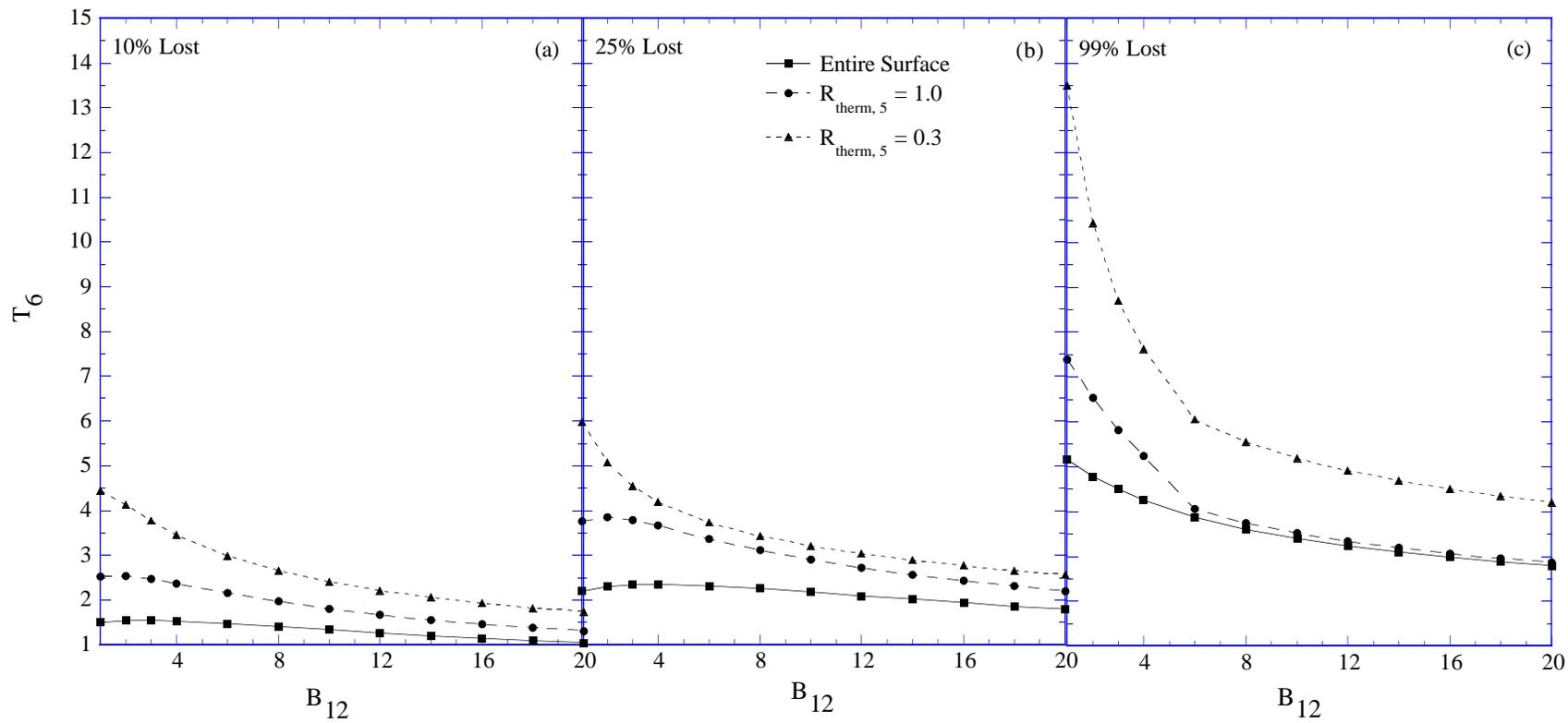

Figure 5